# Correlation-Induced Band Competition in SrTiO$_3$/LaAlO$_3$

Eran Maniv[1], Yoram Dagan[1], and Moshe Goldstein[1]

[1]Raymond and Beverly Sackler School of Physics and Astronomy, Tel Aviv 6997801, Israel

**ABSTRACT**

The oxide interface SrTiO$_3$/LaAlO$_3$ supports a 2D electron liquid displaying superconductivity and magnetism, while allowing for a continuous control of the electron density using a gate. Our recent measurements have shown a similar surprising nonmonotonic behavior as function of the gate voltage (carrier density) of three quantities: the superconducting critical temperature and field, the inverse Hall coefficient, and the frequency of quantum oscillations. While the total density has to be monotonic as function of gate, the last result indicates that one of the involved bands has a nonmontonic occupancy as function of the chemical potential. We show how electronic interactions can lead to such an effect, by creating a competition between the involved bands and making their structure non-rigid, and thus account for all these effects. Adding Fock terms to our previous Hartree treatment makes this scenario even more generic.

**INTRODUCTION**

Understanding correlated oxide interfaces is of high importance from both fundamental science perspective and the prospect of future applications [1]. A prime example is the (100) interface between the insulators SrTiO$_3$ and LaAlO$_3$ (STO/LAO) [2]: when the LAO layer is at least four unit cells thick, a 2D conducting layer appears between them [3]. The density of electrons in this layers can be easily tuned by a back [3] or top gate [4], similarly to semiconductor 2D electron gas; but, in marked difference from semiconductors, the STO/LAO interface shows correlation effects, such as superconductivity [5,6] and magnetism [7,8].

Naively, one would expect that upon increase of the applied gate voltage $V_g$, and therefore of the carrier density at the interface, the properties of the system would evolve in a monotonic fashion, and it would behave as a progressively better conductor. Surprisingly, this turns out not to be the case. For example, the superconducting critical temperature $T_c$ and field $H_{c2}$ first grow but then decrease as function of $V_g$ [6,9,10] leading to a dome shape, reminiscent of high-$T_c$ and other exotic superconductors [11] (see Figure 1). Moreover, the inverse Hall coefficient (which in a simple single-band scenario corresponds to the carrier density) is also non-monotonic, and shows a maximum at roughly the same value of $V_g$ as $T_c$ and $H_c$. These observations by themselves are hard to interpret, since in STO/LAO multiple bands are expected to play a role [13]. However, recently we have measured Shubnikov-de Hass (SdH) oscillations at various values $V_g$ [10], which should directly probe the 2D area enclosed by the Fermi line, and therefore the electronic density in the band which is mobile enough to contribute to the quantum oscillations. Strikingly, the density extracted from the SdH measurement is also non-monotonic with a peak at similar $V_g$ value to the previous phenomena. On the other hand, this density is about an order of magnitude smaller than that corresponding to the inverse Hall coefficient (Figure 2).

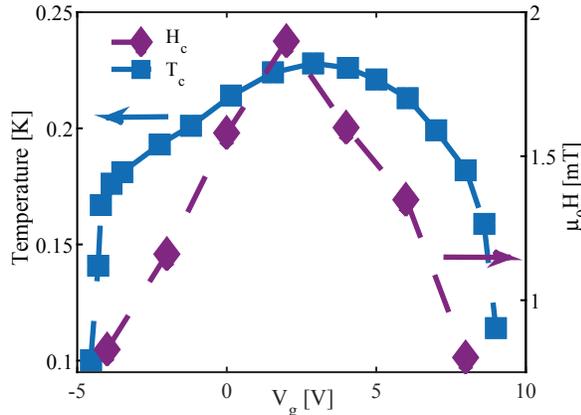

**Figure 1. Nonmonotonic behavior of the superconducting properties (Experiment).** The sample measured has 6 unit cells of LAO grown on STO as described in ref. [12]. The Hall-bar dimensions are 9×3 μm$^2$. Left axis (indicated by an arrow): superconducting critical temperature $T_C$, defined as the temperature where the resistance drops by 50% with respect to the normal state, vs. back gate voltage ($V_g$) [squares]. Right axis (indicated by an arrow): superconducting critical magnetic field $H_{C2}$, defined as the magnetic field where the resistance drops by 50% with respect to the normal state, vs. $V_g$ at T=60mK [diamonds]. Both $T_C$ and $H_C$ exhibit a similar behavior with a maximum at the same $V_g$.

It should be stressed that, as dictated by the laws of thermodynamics, the overall charge density has to be monotonic as function of $V_g$, even when accounting for the nonlinear dielectric behavior of STO [14]. Therefore, the above-mentioned phenomena imply that the band structure is not rigid, and consequently the distribution of carriers between the different bands is changing nonmonotonically (while the overall density is monotonic).

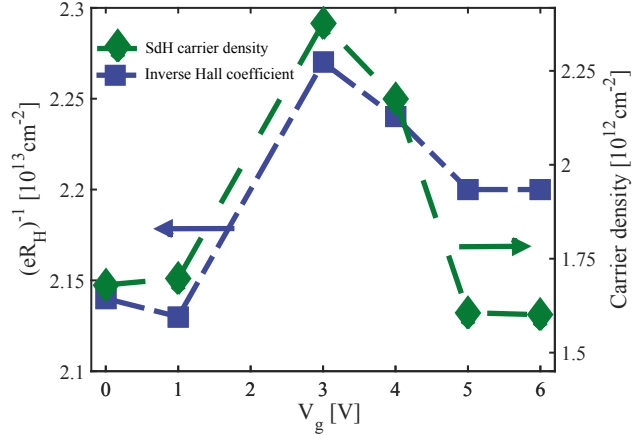

**Figure 2. Nonmonotonic behavior of the transport properties (Experiment).** Left axis (indicated by an arrow): the inverse of the Hall coefficient, inferred from a linear fit to the Hall measurements up to 2T vs. $V_g$ [squares]. Right axis (indicated by an arrow): the carrier density extracted from SdH measurements versus $V_g$ [diamonds]. Both the inverse Hall coefficient and the SdH carrier density exhibit a nonmonotonic behavior with a maximum at about the same $V_g$ as $T_c$ and $H_c$ (see Figure. 1).

In a previous work [10] we have explored this idea. We employed a tight-binding model (introduced previously in [9]), and added to it a new element: local Hubbard interaction, which we treated in the Hartree approximation, and successfully accounted for all these observations. Here we improve the calculation by incorporating the Fock terms. As we discuss below, this leads to an effective enhancement of the spin-orbit coupling in the material, and makes the nonmontonic behavior appear in a much wider range of the parameter space, in particular for lower interaction strengths.

**MAIN THEORETICAL IDEA**

Before going into the details of the calculation, let us outline the mechanism we put forward for explaining the non-monotonic phenomena.

Ab-initio calculations [15] showed that conducting electrons reside on the titanium $t_{2g}$ bands. The degeneracy of the three constituent bulk bands, $d_{xy}$, $d_{yz}$ and $d_{xz}$, is lifted by the confinement [16]: The $d_{xy}$ band, which is heavier in the confinement direction (and lighter in the 2D plane) has a lower energy than the $d_{xz}/d_{yz}$ bands, which are lighter in the confinement direction (and heavier in the 2D plane). Joshua *et al.* [9] added to these basic ingredients atomic spin-orbit coupling, which intermixes the orbitals and removes the crossing between the two upper bands. In this noninteracting picture the band structure is independent of the chemical potential. However, such a model cannot account for the nonmonotonicities observed in the experiment, especially the SdH measurement, which prompts us to incorporate on-site Hubbard interaction, whose strength we denote by U. With this the band structure becomes non-rigid.

Upon increasing the chemical potential from a low value, the first band (which in this regime is almost purely of light $d_{xy}$ character) is gradually filled, until a certain point where the second band starts getting populated. If U is large enough, a competition between the bands emerges: when one of them can gets populated, the energy of the other effectively increases and it empties. In such a situation it is preferred to populate the heavier band, which has higher density of states, and therefore lower single-particle energy. When the second band just starts getting populated it mainly has a heavy $d_{xz}/d_{yz}$ character and is thus favored. However, due to the band mixing the first band quickly becomes heavy and the second light. The system therefore lowers its energy by transferring electrons between them, making the population of the second band nonmonotonic as function of the chemical potential.

With this we can account for all the experimental observations. Heavy bands contribute very little to quantum oscillations. Therefore, the measured SdH frequency should correspond to the density of electrons in the second band (at low values of the chemical potential the first band is light; however, as ab-initio calculations indicate [17], in that region the relevant Ti $d_{xy}$ orbitals reside in the vicinity of the interface, and are therefore more prone to disorder there, and less visible in SdH), which is both smaller than the total density and nonmonotonic, in accordance with the measurement. The inverse Hall coefficient gets contributions from both bands and corresponds to a higher density value, but also shows non-monotonic behavior due to the different mobilities of the two involved bands. Finally, the second more mobile band should dominate superconductivity; the nonmonotonic behavior of its occupancy, and therefore of its density of states, would make the superconducting $T_c$ and $H_c$ nonmonotonic as well.

Let us note that the correlation between SdH and $T_c/H_c$ data shows that the second band appears well below the maximum in $T_c/H_c$, so the superconducting dome maximum does not correspond to a Lifshitz point, in contrast to the interpretation in [9]. Rather, we conjecture that the Lifshitz point (where the second band starts becoming populated) corresponds to the onset of superconductivity. The dome results from the nonmonotonicity of this band's properties as function of $V_g$.

After laying out the main qualitative points in our theory, let us turn to quantitative calculation supporting this scenario.

**MODEL AND RESULTS**

As the single-particle part of our model we employ the Hamiltonian and parameter values introduced by Jushua *et al.* [9,18], based on ab-initio [15] and ARPES data [16]. At every point **k** in the Brillouin zone (BZ) the Hamiltonian is a 6x6 matrix in the tensor-product basis of the three orbitals j=xy,xz,yz times the spin s=↑,↓. Upon diagonalization it gives rise to three bands, m=1,2,3, each of which is almost two-fold degenerate at each **k** point, corresponding to an index α=1,2. Its spin independent part assumes the form

$$H_0(\mathbf{k}) = \begin{pmatrix} \varepsilon_{xy}(\mathbf{k}) - \Delta_E & i\Delta_Z \sin(k_y a) & i\Delta_Z \sin(k_x a) \\ -i\Delta_Z \sin(k_y a) & \varepsilon_{xz}(\mathbf{k}) & t_d \sin(k_x a)\sin(k_y a) \\ -i\Delta_Z \sin(k_x a) & t_d \sin(k_x a)\sin(k_y a) & \varepsilon_{yz}(\mathbf{k}) \end{pmatrix} \otimes I_2, \qquad (1)$$

where

$$\varepsilon_{xy}(\mathbf{k}) = t_l[2 - \cos(k_x a) - \cos(k_y a)],$$

$$\varepsilon_{xz}(\mathbf{k}) = t_l[1 - \cos(k_x a)] + t_h[1 - \cos(k_y a)],$$

$$\varepsilon_{yz}(\mathbf{k}) = t_h[1 - \cos(k_x a)] + t_l[1 - \cos(k_y a)],$$
(2)

with the $t_l$=875meV, $t_h$=$t_d$=40meV being, respectively, the nearest-neighbor light and heavy, as well as next-nearest-neighbor tight-binding overlap integrals, with a=3.905Å being the lattice constant. $\Delta_E$=47meV is the $d_{xy}$ to $d_{xz}/d_{yz}$ splitting due to the confinement. $\Delta_Z$=20meV quantifies the breaking of mirror symmetry about the xy plane in the heterostructure.

The atomic spin-orbit coupling involves the Pauli matrices $\sigma_x$, $\sigma_y$, and $\sigma_z$,

$$H_{SO} = \frac{i\Delta_{SO}}{2}\begin{pmatrix} 0 & \sigma_x & -\sigma_y \\ -\sigma_x & 0 & \sigma_z \\ \sigma_y & -\sigma_z & 0 \end{pmatrix},$$
(3)

where the spin-orbit coupling strength is $\Delta_{SO}$=9meV. We have previously shown the Rashba spin-orbit coupling to have negligible effects [10].

In the Hartree-Fock approximation, the on-site Hubbard interaction U modifies the single-particle Hamiltonian by the addition of the term:

$$H_{int} = U\begin{pmatrix} (N_{xy} + 4N_z)I_2 & -iN_{SO}^{xy}\sigma_x & iN_{SO}^{xy}\sigma_y \\ iN_{SO}^{xy}\sigma_x & (2N_{xy} + 3N_z)I_2 & -iN_{SO}^{z}\sigma_z \\ -iN_{SO}^{xy}\sigma_y & iN_{SO}^{z}\sigma_z & (2N_{xy} + 3N_z)I_2 \end{pmatrix},$$
(4)

which is the most general one allowed by the symmetries of the system (time reversal, $C_4$ rotation, and reflection through the xz and yz mirror planes; those are broken by magnetism [7,8] and by the tetragonal structural transition of STO [19], but the corresponding energy scales are too small to appreciably affect our results). The diagonal part represents the Hartree contribution, which includes the average occupancies of the j=xy,xz,yz orbitals,

$$N_j = \frac{1}{2}\sum_{m,\alpha,s}\int_{BZ}\frac{a^2 d^2\mathbf{k}}{(2\pi)^2}|\psi_{m,\alpha}(\mathbf{k};j,s)|^2 f[\varepsilon_{m,\alpha}(\mathbf{k})],$$
(5)

These depend on the band eigenfunctions and eigenenergies, with f(ε) being the fermi function. The symmetries of the systems dictate $N_{xz}$=$N_{yz}$=$N_z$. The off-diagonal parts in (4) correspond to the Fock terms, which renormalize (enhance) the spin-orbit coupling $\Delta_{SO}$. They include the following two coefficients,

$$N_{SO}^{xy} = \text{Im}\sum_{m,\alpha,s}\int_{BZ}\frac{a^2 d^2\mathbf{k}}{(2\pi)^2}\psi_{m,\alpha}(\mathbf{k};xy,\uparrow)\psi_{m,\alpha}^*(\mathbf{k};xz,\downarrow)f[\varepsilon_{m,\alpha}(\mathbf{k})]$$

$$= -\text{Re}\sum_{m,\alpha,s}\int_{BZ}\frac{a^2 d^2\mathbf{k}}{(2\pi)^2}\psi_{m,\alpha}(\mathbf{k};xy,\uparrow)\psi_{m,\alpha}^*(\mathbf{k};yz,\downarrow)f[\varepsilon_{m,\alpha}(\mathbf{k})],$$
(6)

$$N_{SO}^{z} = \text{Im}\sum_{m,\alpha,s}\int_{BZ}\frac{a^2 d^2\mathbf{k}}{(2\pi)^2}\psi_{m,\alpha}(\mathbf{k};xz,\uparrow)\psi_{m,\alpha}^*(\mathbf{k};yz,\uparrow)f[\varepsilon_{m,\alpha}(\mathbf{k})].$$
(7)

The Hartree-Fock coefficients are determined self consistently. After extracting them one may find all the measureable quantities. The band occupancies are:

$$n_m = \sum_{\alpha}\int_{BZ}\frac{d^2\mathbf{k}}{(2\pi)^2}f[\varepsilon_{m,\alpha}(\mathbf{k})].$$
(8)

Their derivative with respect to the chemical potential m (keeping the Hartree-Fock parameters fixed) determines the corresponding densities of states. The longitudinal and Hall conductivities (the later in the presence of a weak perpendicular magnetic field B) are given by [20]:

$$\sigma_L = -\frac{e^2}{2} \sum_{m,\alpha} \int_{BZ} \frac{d^2\mathbf{k}}{(2\pi)^2} \left\{ [v_{m,\alpha}^x(\mathbf{k})]^2 + [v_{m,\alpha}^y(\mathbf{k})]^2 \right\} \tau_{m,\alpha}(k) f'[\varepsilon_{m,\alpha}(\mathbf{k})], \quad (9)$$

$$\sigma_H = -e^3 B \sum_{m,\alpha} \int_{BZ} \frac{d^2\mathbf{k}}{(2\pi)^2} v_{m,\alpha}^x(\mathbf{k}) \tau_{m,\alpha}(\mathbf{k}) f'[\varepsilon_{m,\alpha}(\mathbf{k})] \left[ v_{m,\alpha}^y(\mathbf{k}) \frac{1}{\hbar} \frac{\partial}{\partial k_x} - v_{m,\alpha}^x(\mathbf{k}) \frac{1}{\hbar} \frac{\partial}{\partial k_y} \right] v_{m,\alpha}^y(\mathbf{k}) \tau_{m,\alpha}(\mathbf{k}), \quad (10)$$

where $\mathbf{v}_{m,\alpha}(\mathbf{k}) = \partial \varepsilon_{m,\alpha}(\mathbf{k})/\hbar \partial \mathbf{k}$ is the velocity vector, and the transport lifetime is taken for simplicity as a weighted average of the $d_{xy}$ and $d_{xz}/d_{yz}$ lifetimes $\tau_{xy}$ and $\tau_z$, respectively:

$$\tau_{m,\alpha}(\mathbf{k}) = \sum_s |\psi_{m,\alpha}(\mathbf{k}; xy, s)|^2 \tau_{xy} + \left[ |\psi_{m,\alpha}(\mathbf{k}; xz, s)|^2 + |\psi_{m,\alpha}(\mathbf{k}; yz, s)|^2 \right] \tau_z. \quad (11)$$

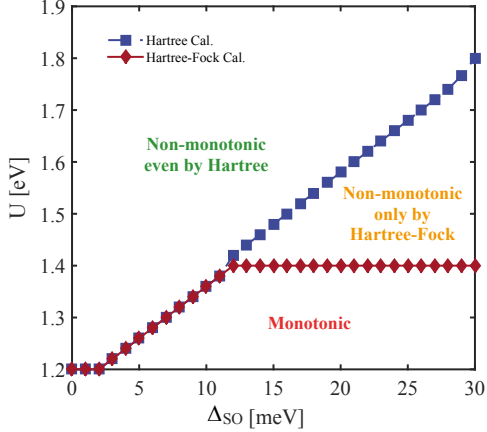

**Figure 3. "Phase diagram" (Theory).** Regions of monotonic and nonmontonic behavior of the second band population vs. chemical potential in the Hubbard U-spin orbit coupling plane. The Fock terms make the nonmonotonic behavior more generic.

In a previous work [10] we have studied this model without including the Fock terms, Equations (6) and (7). As shown in Figure 3, the Fock terms make the nonmonotonic behavior of the second band population more generic, as they allow it to appear for lower U values. Accounting for the experimental data is more demanding: the non-monotonicity of the second band should be significant, and the third band should be pushed to higher energies. By enhancing the spin-orbit coupling, the Fock terms make these requirements easier to achieve. We can thus reproduce the behavior of density of the second band (and hence the SdH oscillations; Figure 4), as well as the inverse Hall coefficient and the second band density of states (which affects superconductivity; Figure 5) with a mild value of the atomic spin-orbit coupling $\Delta_{SO}$ (as appropriate for the relatively light Ti [9,18]), and for U=2.2eV, close to the value extracted from STM measurements [21]. The corresponding chemical potential range agrees with the range of $V_g$ values in Figures 1 and 2, based on our estimates for the capacitance of the sample [10] (Note that, due to the polarity of LAO, $V_g=0$ does not correspond to $\mu=0$).

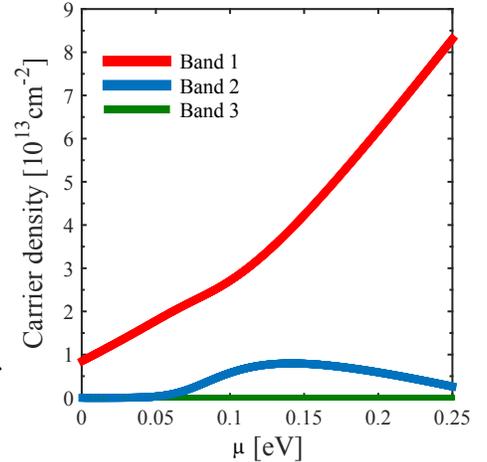

**Figure 4. Nonmonotonic behavior of the band densities (Theory).** Carrier densities of the three bands are plotted as function of the chemical potential $\mu$ calculated using the Hartree Fock approximation for U=2.2eV and $\Delta_{SO}$=9meV. The second band population (corresponding to the SdH frequency) is nonmonotonic and much smaller than that of the first band.

## CONCLUSIONS

We have shown how incorporating interaction effects can lead to competition between bands and modifications of the band structure of STO/LAO, and therefore to a nonmonotonic behavior of all the quantities related to the second band (density, Hall effect, and density of states), in agreement with our measurements (SdH, Hall, and superconducting dome, respectively). Including the Fock terms in our calculation makes this scenario more generic.

Recently, Smink *et al.* [22] have measured the influence of both top and bottom gates, and found them to have similar behavior. This excludes the particular potential profile created by a back gate as an explanation for the nonmonotonic behavior. Furthermore, they accompanied their data with a theoretical calculation which includes a modeling of the confinement using the Schrodinger-Poisson equations, but on the other hand contains neither spin-orbit coupling nor the Fock terms. This leads to qualitatively similar results to our previous Hartree calculations [10]. In the future it would be interesting to study the interplay between the Poisson-Schrodinger description of the confinement and the spin orbit and Fock couplings, as well as to study superconductivity in an appropriate multi-band model.

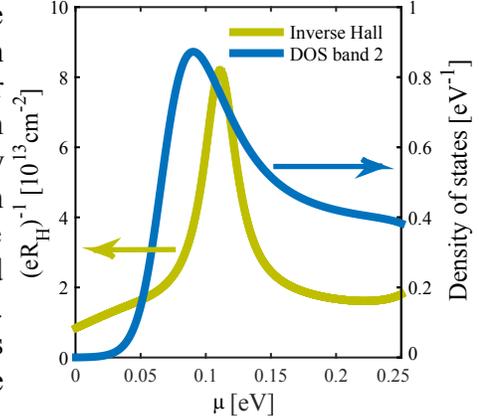

**Figure 5. Nonmonotonic behavior of the measureable properties (Theory).** Left axis (indicated by an arrow): the inverse Hall coefficient vs. $\mu$. Right axis (indicated by an arrow): the second band density of states per unit cell vs. $\mu$. Both are calculated using the Hartree-Fock approximation for U=2.2eV, $\Delta_{SO}$=9meV, and $\tau_{xy}/\tau_z$=8 (making the second band more mobile in the relevant region, in accordance with its visibility in SdH). Both display similar nonmonotonic behavior to the second band occupancy (Figure 4).

## ACKNOWLEDGMENTS

We thank Yotam Gigi for preliminary computer simulations. This work was partially supported by the Israeli Science Foundation under grants 569/13 and 227/15, by the Ministry of Science and Technology under contracts 3-11875 and 3-12419, and by the Pazy foundation under contract 268/17.